\documentclass[iop, apj]{emulateapj}

\newcommand{\kms}{\rm km~s^{-1}}
\newcommand{\kmsmpc}{\rm km~s^{-1}~Mpc^{-1}}
\newcommand{\dn}{D_{n}4000}

\slugcomment{\bf Last updated:\today}

\usepackage{color}
\usepackage{multirow}
\usepackage{amsmath}  
\usepackage[breaklinks,colorlinks,
   urlcolor=blue,citecolor=blue,linkcolor=blue]{hyperref}

\begin{document}

\title{The Velocity Dispersion Function for Quiescent Galaxies
       in the Local Universe}

\author{Jubee Sohn$^{1}$,
	    H. Jabran Zahid$^{1}$,
        Margaret J. Geller$^{1}$} 

\affil{$^{1}$ Smithsonian Astrophysical Observatory, 60 Garden Street, Cambridge, MA 02138, USA}

\begin{abstract}
We investigate the distribution of central velocity dispersions 
 for quiescent galaxies in the SDSS at $0.03 \leq z \leq 0.10$. 
To construct the field velocity dispersion function (VDF), 
 we construct a velocity dispersion complete sample 
 of quiescent galaxies with $\dn > 1.5$. 
The sample consists of galaxies with central velocity dispersion 
 larger than the velocity dispersion completeness limit 
 of the SDSS survey. 
Our VDF measurement is consistent with 
 previous field VDFs for $\sigma > 200~\kms$. 
In contrast with previous results, 
 the VDF does not decline significantly for 
 $\sigma < 200 ~\kms$.  
The field and the similarly constructed cluster VDFs are remarkably flat 
 at low velocity dispersion ($\sigma < 250~\kms$). 
The cluster VDF exceeds the field for $\sigma > 250~\kms$
 providing a measure of the relatively larger 
 number of massive subhalos in clusters. 
The VDF is a probe of the dark matter halo distribution
 because the measured central velocity dispersion may be
 directly proportional to the dark matter velocity dispersion. 
Thus the VDF provides a potentially powerful test of simulations 
 for models of structure formation. 
\end{abstract}

\section{INTRODUCTION}
The stellar velocity dispersion of galaxy is 
 a fundamental observable that is correlated with 
 other basic properties of galaxies 
 including luminosity ($L-\sigma^{4}$ relation, \citealp{Faber76}) and 
 size and surface brightness (the fundamental plane, \citealp{Djorgovski87, Dressler87}). 
The velocity dispersion, often measured within the central region of a galaxy,
 reflects the dynamics of its stellar population.
The velocity dispersion for a quiescent galaxy is particularly interesting 
 because it is related to the mass of the galaxy through
 the virial theorem (e.g. \citealp{Faber76, Bezanson11, Zahid17}). 

The central velocity dispersion (velocity dispersion hereafter) 
 is a robust spectroscopic measure 
 insensitive to the photometric biases 
 that impact measurements of other fundamental observables 
 including luminosity and stellar mass \citep{Bernardi13}.
Thus, the velocity dispersion of the stellar population 
 may be the best observable connect galaxies 
 directly to their dark matter (DM) halos
 \citep{Wake12a, Wake12b, Bogdan15, Zahid16}. 
\citet{Zahid16} show that 
 there is a scaling between central velocity dispersion and stellar mass, 
 $\sigma \varpropto M_{*}^{0.3}$. 
This scaling is consistent with the scaling 
 between the velocity dispersion of DM halo and total halo mass, 
 $\sigma_{DM} \varpropto M_{halo}^{0.28 - 0.33}$, 
 measured from N-body simulations \citep{Evrard08, Posti14}.  
Based on the consistency of these scaling relations, 
 \citet{Zahid16} suggest that 
 the measured central velocity dispersion is proportional to 
 the velocity dispersion of DM halo.  
Several studies also show that 
 the central stellar velocity dispersion of elliptical galaxies 
 is essentially identical to the velocity dispersion 
 estimated from the best-fit strong lensing models 
 (e.g. \citealp{Treu06, Grillo08, Auger10}). 
These strong lensing studies underscore 
 the fundamental importance of the central velocity dispersions.

The velocity dispersion distribution, 
 i.e., the velocity dispersion function (VDF hereafter), 
 has been measured for general field galaxies based on the SDSS and BOSS galaxy surveys
 \citep{Sheth03, Mitchell05, Choi07, Bernardi10, Montero-Dorta17}.
These VDFs measured from the general galaxy population, 
 which we refer to as field VDFs, 
 have similar shapes at $\sigma > 150~\kms$, 
 but they differ substantially for $\sigma \leq 150~\kms$.
\citet{Choi07} attribute the discrepancy at the low velocity dispersion to 
 differences in sample selection 
 including the definition of quiescent galaxies. 
Comparison among these previous field VDFs suggests that 
 sample selection is critical for determining the shape of the VDF. 
 
VDFs of cluster galaxies also differ from the field VDFs
 \citep{Munari16, Sohn17}. 
\citet{Sohn17} compare VDFs for the massive clusters Coma and A2029 with previous field VDFs.
They show that the cluster VDFs significantly exceed the field VDFs 
 at high ($\sigma > 250~\kms$) and low ($\sigma < 150~\kms$)
 velocity dispersion. 
They suggest that the relatively large abundance of massive cluster members
 with large velocity dispersion results in the excess for $\sigma > 250~\kms$. 
The cluster sample of \citet{Sohn17} 
 differs from the previous field samples 
 in the definition of quiescent galaxies and 
 in the corrections to velocity dispersion. 
These differences may be responsible for the difference in the VDFs for $\sigma < 150~\kms$.
Here, we explore this discrepancy by selecting a field sample 
 in a way that is essentially identical to the cluster sample. 
 
Statistical analyses of galaxy properties like the VDF require samples 
 that cover the complete distribution of the galaxy properties. 
For example, luminosity functions are only 
 complete to the absolute magnitude limit of a volume limited sample. 
Likewise, stellar mass functions can be measured from a sample that 
 completely surveys the range of stellar masses
 \citep{Fontana06, Perez-Gonzalez08, Pozzetti10, Weigel16}. 
Stellar mass complete samples have been based on empirically derived stellar mass completeness limits 
 as a function of redshift \citep{Pozzetti10, Weigel16}. 
A robust VDF measurement also requires a sample that
 includes the complete range of the velocity dispersions at each absolute magnitude. 
Here we derive a robust measurement of the VDF by empirically determining the velocity dispersion 
 completeness limit. 
 
We examine the VDF for the general field based on the extensive SDSS spectroscopic survey data. 
In contrast with previous approaches, we construct 
 a velocity dispersion complete sample from the SDSS survey. 
To compare our result directly with the cluster VDFs of \citet{Sohn17}, 
 we select the field sample according to the prescription of \citet{Sohn17}. 
Comparison between the field and cluster VDFs offers 
 potential constraints on structure formation models and 
 may be particularly interesting for understanding the formation and evolution of galaxies in the cluster environment. 

We describe the data in Section 2. 
We explain the construction of the velocity dispersion complete sample in Section 3.
Section 4 describes the method we use to measure the VDF. 
We discuss the results in Section 5 and we conclude in Section 6. 
Throughout the paper, 
 we adopt the standard $\Lambda$CDM cosmology of 
 $H_{0} = 70~\kmsmpc$, $\Omega_{m} = 0.3$, and $\Omega_{\Lambda} = 0.7$. 

\section{DATA}

Our primary goals are 
 1) measuring the velocity dispersion distribution function
 for field galaxies from a complete sample of galaxy central velocity dispersions and 
 2) comparing the VDF
 with previous results for the field and clusters. 
Here, we describe the galaxy sample, the photometric data (Section \ref{phot})
 and the spectroscopic properties of the sample including
 the galaxy central velocity dispersion (velocity dispersion hereafter, Section \ref{sigma}) and 
 the $\dn$ index (Section \ref{dn4000}).
For a fair comparison with the cluster sample, 
 we determine the field galaxy sample in the same way we define the cluster population. 
We construct a dataset that includes 
 the entire range of galaxy velocity dispersions at every K-corrected absolute magnitude
 (see Section \ref{sigsam}). 
We call this dataset the velocity dispersion complete sample. 

\subsection{Photometric Data}\label{phot}
We use the Main Galaxy Sample \citep{Strauss02} from the Sloan Digital Sky Survey (SDSS)
 Data Release 12 \citep{Alam15}. 
The Main Galaxy Sample is a magnitude limited sample 
 with $r_{\rm petro} < 17.77$ and $z \lesssim 0.3$. 
The SDSS spectroscopic survey is $\sim 95\%$ complete.
The spectra cover $3500 - 9000$ {\rm \AA}
 with a resolution of $R \sim 1500$ at 5000 {\rm \AA}.
We use galaxies in the Main Galaxy Sample in the redshift range $0.01 < z < 0.10$. 
We apply the lower redshift limit to minimize 
 the influence of peculiar motions. 

We K-correct the r-band magnitude
 to derive luminosities of galaxies 
 over a wide redshift range. 
We use the $z=0$ K-correction from 
 the NYU Value Added Galaxy Catalog \citep{Blanton05}. 
We correct the SDSS spectroscopic limit 
 by applying a median K-correction as a function of redshift 
 (see Figure 2 in \citealp{Zahid17}). 
We refer to the K-corrected r-band absolute magnitude as $M_{r}$. 

\subsection{Velocity dispersion}\label{sigma}
We adopt the velocity dispersions measured by the Portsmouth reduction \citep{Thomas13}. 
The Portsmouth velocity dispersions are measured from SDSS spectra
 using the Penalized Pixel-Fitting (pPXF) code \citep{Cappellari04}. 
In this procedure, stellar population templates from \citet{Maraston11} 
 which is based on the MILES stellar library \citep{Sanchez06}
 are converted to SDSS resolution. 
By comparing the SDSS spectra and the templates, 
 the best-fit velocity dispersion is derived for each galaxy.
 
We compare the Portsmouth velocity dispersions
 for galaxies in the field sample with 
 velocity dispersions for cluster members 
 from the same reduction (Section \ref{sec_cl}). 
\citet{Sohn17} measure the velocity dispersion function for 
 Coma using velocity dispersion based on the Portsmouth reduction and 
 for Abell 2029 based on the velocity dispersion 
 from observations with Hectospec on MMT \citep{Fabricant05}, respectively.
Because the fiber sizes of SDSS spectrograph and Hectospec differ,
 an aperture correction is necessary. 
Following \citet{Sohn17}, 
 the aperture correction is
\begin{equation}
 (\sigma_{\rm SDSS}/\sigma_{\rm Hectospec}) = (R_{\rm SDSS} / R_{\rm Hectospec})^{-0.054 \pm 0.005},
\label{aper}
\end{equation}
 where $R_{\rm SDSS} = 1.\arcsec5$ and $R_{\rm Hectospec} = 0.\arcsec75$.  
This aperture correction is consistent with 
 previous determinations \citep{Cappellari04, Zahid16}.

Coma ($z=0.023$) and A2029 ($z=0.078$) are at different redshifts. 
Thus, the fibers cover 
 different physical apertures for the galaxies in the two clusters. 
Using equation \ref{aper}, 
 \citet{Sohn17} correct the velocity dispersion to a fiducial aperture of 3 kpc.  

To compare the field and cluster VDFs directly, 
 we use a velocity dispersion within 3 kpc following \citet{Sohn17}. 
We apply the aperture correction to 
 the Portsmouth velocity dispersion for the galaxies 
 in the magnitude limited sample. 
The aperture correction is small ($\sim0.01$ dex) and 
 does not significantly impact our results. 
Hereafter, 
 we refer to this velocity dispersion within a 3 kpc aperture as $\sigma$.

\subsection{$\dn$}\label{dn4000}

We use $\dn$ from the MPA/JHU Catalog 
 \footnote{http://www.mpa.mpa-garching.mpg.de/SDSS/DR7/} 
 to identify quiescent galaxies. 
$\dn$ is a spectroscopic measure of the amplitude of the 4000 {\rm \AA} break.
\citet{Balogh99} define $\dn$ as the ratio of the flux in the
 $4000 - 4100$ {\rm \AA} and $3850 - 3950$ {\rm \AA} bands \citep{Balogh99}. 

$\dn$ is a stellar population age indicator 
 \citep{Kauffmann03, Geller14}
 showing a bimodal distribution \citep{Kauffmann03}.
Thus, the index is often used to segregate star-forming and quiescent galaxies
 \citep{Kauffmann03, Mignoli05, Vergani08, Woods10, Zahid16}. 
$\dn$ is a powerful tool 
 because it can be measured at high signal to noise 
 directly from the spectra. 
This index is insensitive to photometric issues including 
 seeing, crowding of galaxies and reddening. 
$\dn$ is also a redshift independent index. 
Thus, it is a relatively clean basis 
 for construction of VDFs in a wide range of environments and at a wide range of redshifts.

We identify quiescent galaxies as objects with $\dn > 1.5$. 
The $\dn$ selection is the same as the one used by \citet{Sohn17}
 who examine VDFs for quiescent galaxies in Coma and A2029.
\citet{Sohn17} use $\dn$ for A2029 members from Hectospec observations. 
The $\dn$ measurements from SDSS and Hectospec for the same objects are within $\sim5\%$,
 a difference that does not affect our analysis \citep{Fabricant13}.
We note that our results are insensitive to the $\dn$ selection. 
If we select quiescent galaxies with $\dn > 1.4$ or $\dn > 1.6$,
 the velocity dispersion range of the samples varies slightly, 
 but the shapes of the VDFs measured from the samples are indistinguishable.

\section{CONSTRUCTING A VELOCITY DISPERSION COMPLETE SAMPLE}\label{sigsam}

A complete sample is key to measuring
 the statistical distribution of any galaxy property.
For example, 
 the galaxy luminosity function
 may be derived from a volume limited sample 
 (e.g. \citealp{Norberg01, Croton05}). 
Conventional volume limited samples contain 
 every galaxy within a given volume  
 brighter than the absolute magnitude limit,
 i.e. any galaxy in the sample would be observable
 throughout the volume surveyed.  
Statistical distributions of the properties of galaxies 
 such as stellar mass and velocity dispersion
 can also be examined based on volume limited samples (e.g. \citealp{Choi07}). 
However, these measurements may be biased or incomplete
 because the volume limited sample is only complete for 
 a range in absolute magnitude, not for a range in either the stellar mass 
 and/or the velocity dispersion \citep{Zahid17}.
These properties of galaxies are correlated with galaxy luminosity, 
 but there is substantial scatter around each relation.
In other words, at each absolute magnitude, 
 the volume limited samples may not include 
 the full range of stellar mass or velocity dispersion.  
 
Here, we describe the construction of a sample 
 for measuring the complete VDF from a magnitude limited sample. 
The magnitude limited sample
 includes galaxies with $r < 17.77$, $0.01 \leq z \leq 0.10$, and $\dn > 1.5$. 
We first demonstrate the incompleteness of 
 a conventional volume limited sample 
 in terms of $\sigma$ (Section \ref{volsam});  
 the full range of $\sigma$ may not be sampled at every magnitude
 because of the large scatter in the $\sigma$ distribution at fixed absolute magnitude. 
We derive $\sigma_{lim} (z)$, 
 the limit where the sample includes the full range of $\sigma$ 
 for every absolute magnitude
 in sample covering the range $0.03 \leq z \leq 0.10$ (Section \ref{sigfun}). 
 
We begin by examining the scatter in $\sigma$ at a fixed absolute magnitude 
 in a volume limited sample. 
We identify $\sigma_{lim}$, 
 the $\sigma$ complete limit where the volume limited sample is complete 
 for $\sigma > \sigma_{lim}$ 
 despite the scatter in the $\sigma$-to-light ratio.
This $\sigma_{lim}$ is the $\sigma$ completeness limit at the maximum redshift ($z_{max}$)
 of the volume limited sample. 
We then determine $\sigma_{lim}(z)$ by repeating 
 the $\sigma_{lim}$ determination for 
 a series of volume limited samples with different $z_{max}$. 
Finally, we construct a velocity dispersion complete sample 
 (hereafter $\sigma$ complete sample)
 consisting of galaxies with $\sigma > \sigma_{lim}(z)$ at each z. 
Table \ref{term} lists the terminology we use and 
 Table \ref{sample} summarizes the various sample we construct. 
 
\begin{deluxetable*}{cl}
\tablecolumns{2}
\tabletypesize{\scriptsize}
\tablewidth{0pt}
\tablecaption{Terminology}
\tablehead{\colhead{Symbol} & \colhead{Quantity/Variable}}
\startdata
$M_{r}$ & K-corrected r-band magnitude \\
$\sigma$ & Central velocity dispersion \\
         & (aperture corrected to a 3 kpc fiducial aperture) \\
$\Delta \sigma$    & Velocity dispersion uncertainty \\
$M_{r, lim}$       & Magnitude limit of the volume limited sample \\
$\sigma_{lim}$     & Central velocity dispersion completeness limit of a volume limited sample \\
$\sigma_{lim} (z)$ & Central velocity dispersion completeness limit as a function of redshift \\
\hline
$N_{gal}$              & Total number of galaxies in the sample \\
$N_{bin}$              & Number of $\sigma$ bins \\
$\sigma_{j}$           & Central $\sigma$ of the $j$th $\sigma$ bin \\
$\sigma_{lim, i}$      & Minimum $\sigma$ where $i$th galaxy can still be found in the sample \\
$\sigma_{f}$           & Fiducial $\sigma$ for deriving $g$ (equation (6)), $\log \sigma_{f} = 1.0$ \\
$\beta$                & Constant for deriving $g$ (equation (6)), $\beta = 1.5$ \\
$\Omega_{\rm survey}$  & Solid angle of the SDSS survey, $\Omega_{\rm survey} = 8032$ deg$^{2}$ \\
$\Omega_{\rm all~sky}$ & Solid angle of the full sky, $\Omega_{\rm all~sky} = 40253$ deg$^{2}$ \\
$d_{c} (z)$            & Comoving distance at redshift z \\
$z'_{min}$             & Lower redshift limit of the $\sigma$ complete sample, $z'_{min} = 0.03$ \\
$z'_{max}$             & Upper redshift limit of the $\sigma$ complete sample, $z'_{max} = 0.10$
\enddata
\label{term}
\end{deluxetable*}

\begin{deluxetable*}{ll}
\tablecolumns{2}
\tabletypesize{\scriptsize}
\tablewidth{0pt}
\tablecaption{Sample Definition}
\tablehead{\colhead{Identification} & \colhead{Selection}}
\startdata
Magnitude limited sample & $0.01 \leq z \leq 0.10$, $\dn > 1.5$ \\
VL1 & $0.01 \leq z \leq 0.09$, $\dn > 1.5$, $M_{r} < M_{r, lim} = -20.49$ \\
VL2 & $0.03 \leq z \leq 0.09$, $\dn > 1.5$, $M_{r} < M_{r, lim} = -20.49$ \\
$\sigma$ complete sample & $0.03 \leq z \leq 0.10$, $\dn > 1.5$, $\sigma > \sigma_{lim} (z)$
\enddata
\label{sample}
\end{deluxetable*}

\subsection{Velocity Dispersion Incompleteness of Volume Limited Samples}\label{volsam}

\begin{figure*}
\centering
\includegraphics[scale=0.65]{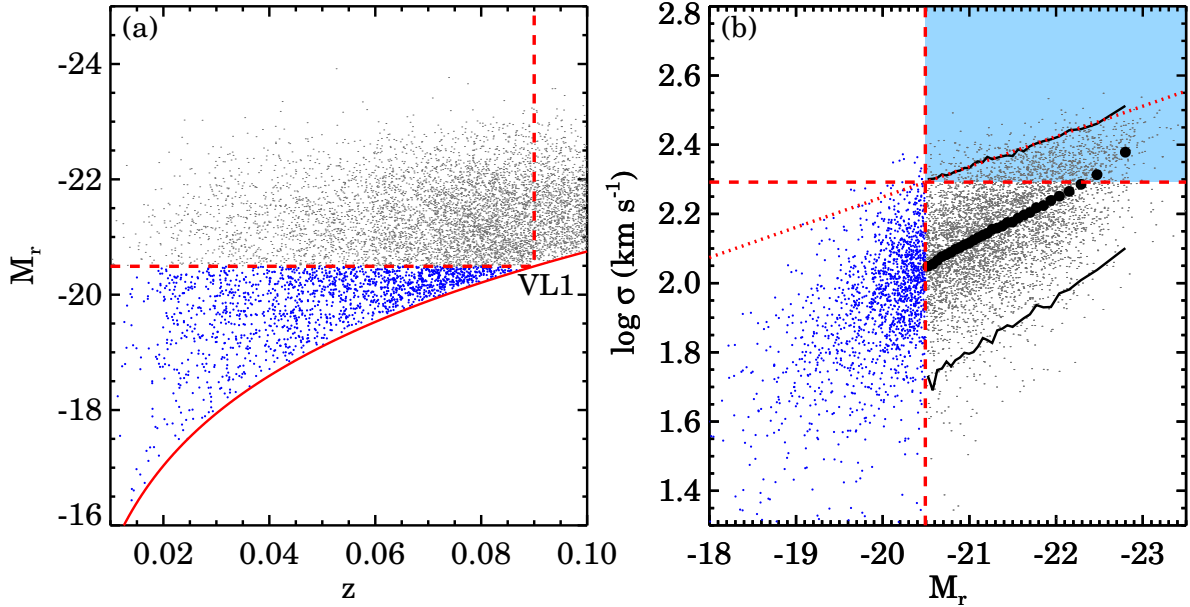}
\caption{(a) K-corrected r-band absolute magnitude, $M_{r}$, as a function of redshift
 for the magnitude limited sample.  
The solid red curve displays the SDSS spectroscopic survey limit ($r = 17.77$). 
The dashed line shows the boundary for volume-limited sample (VL1) 
 with $0.01 < z < 0.09$ and $M_{r} < -20.49$. 
Blue points indicate the objects excluded from VL1. 
(b) $\sigma$ vs. $M_{r}$ for the magnitude limited sample. 
The red vertical dashed line shows the magnitude limit of VL1. 
Black circles and solid curves show the median $\sigma$ and 
 the limits of the central 95\% of the $\sigma$ distribution, respectively. 
The red dotted line is the fit to the upper 95\% limit. 
The $\sigma$ value of the intersection is 
 the $\sigma$ completeness limit ($\sigma_{lim}$, horizontal dashed line)
 at the maximum redshift of VL1, $z_{max} = 0.09$.
For $\sigma < \sigma_{lim}$,  
 a significant fraction of galaxies are 
 fainter than $M_{r, lim}$.
In other words, VL1 only samples the full $\sigma$ distribution at each $M_{r}$
 in the shaded region. 
We display only 5\% of the data for clarity. }
\label{volume}
\end{figure*}

Figure \ref{volume} (a) shows $M_{r}$ 
 as a function of redshift for galaxies 
 in the magnitude limited sample.
The solid line displays the apparent magnitude limit, $r < 17.77$,
 where the SDSS spectroscopic survey is complete
 ($\sim 95\%$, \citealp{Strauss02}). 
A conventional volume limited sample includes
 every galaxy brighter than the $M_{r}$ limit ($M_{r, lim}$) 
 of the sample within the volume limited by the redshift $z_{max}$. 
$M_{r, lim}$ is the $M_{r}$ where the magnitude limited sample 
 is complete at $z_{max}$. 
The dashed lines in Figure \ref{volume} (a) show 
 the boundary of an example volume limited sample (VL1)
 with $0.01 \leq z \leq 0.09$ and $M_{r} \leq M_{r, lim} = -20.49$. 
Within this volume, VL1 contains all galaxies 
 brighter than $M_{r} \leq -20.49$. 
A luminosity function derived from VL1 is then complete to $M_{r} = -20.49$. 
  
Figure \ref{volume} (b) shows 
 $\sigma$ versus $M_{r}$ for galaxies in the magnitude limited sample. 
In general, 
 $\sigma$ increases with luminosity, 
 but the $\sigma$ distribution at a given $M_{r}$ is very broad 
 (see also Figure 14 in \citealp{Sohn17}).  
For example, a galaxy at the limiting magnitude $M_{r} = -20.49$ 
 can have $\sigma$ ranging from $\sim 50~\kms$ to $\sim 200~\kms$. 
Because of the large scatter in $\sigma$, 
 direct conversion from the luminosity limit to a $\sigma$ limit
 is not trivial \citep{Sheth03, Sohn17}. 

A conventional volume limited sample 
 is only complete to the $M_{r}$ limit, not to any $\sigma$ limit. 
The vertical dashed line in Figure \ref{volume} (b) shows 
 the $M_{r}$ limit of VL1. 
Because of the broad $\sigma$ distribution at fixed $M_{r}$,  
 the sample includes many low $\sigma$ galaxies (e.g., $\log \sigma < 2.0$). 
The VL1 sample also excludes galaxies with $\log \sigma > 2.0$
 below the $M_{r}$ limit. 
Blue points in Figure \ref{volume} show the faint galaxies 
 that are excluded from VL1.
The fraction of these excluded objects is significant for $\log \sigma < 2.3$. 
Because of incomplete sampling of these 
 lower luminosity objects,
 the VDF derived directly from VL1
 underestimates the number of objects with 
 $\log \sigma < 2.3$. 

\subsection{Velocity Dispersion Limit as a Function of Redshift}\label{sigfun}

To construct a $\sigma$ complete sample, 
 we derive $\sigma_{lim} (z)$ for the magnitude limited sample. 
The $\sigma$ completeness limit of the magnitude limited sample 
 depends on redshift just as the $M_{r}$ limit does. 
We derive the $\sigma$ completeness limit empirically. 
Several previous studies of the stellar mass functions follow
 a similar approach in constructing a stellar mass complete sample
 \citep{Fontana06, Perez-Gonzalez08, Marchesini09, Pozzetti10, Weigel16}. 

We first derive $\sigma_{lim}$ for a single volume limited sample. 
In Figure \ref{volume} (b), 
 the red solid curves are the empirically determined 
 central 95\% completeness limits for the distribution of $\sigma$
 in VL1 as a function of $M_{r}$. 
The blue dotted line is a linear fit to the upper $\sigma$ limit. 
We determine the intersection of the $M_{r}$ limit (the vertical dashed line)
 and the fit to the upper $\sigma$ limit; 
 the $\sigma$ value of the intersection (the horizontal dashed line) 
 is the $\sigma$ limit of VL1, $\sigma_{lim, VL1}$.
Galaxies fainter than the $M_{r}$ limit of VL1
 rarely ($< 2.5\%$) appear above this $\sigma_{lim, VL1}$.
Thus, VL1 is complete for $\sigma \geq \sigma_{lim, VL1}$.
VL1 is a complete subset of the magnitude limited sample to $z_{max} = 0.09$. 
Therefore, $\sigma_{lim, VL1}$ corresponds to the $\sigma$ completeness limit of the magnitude limited sample 
 at $z_{max} = 0.09$. 

\begin{figure}
\centering
\includegraphics[scale=0.49]{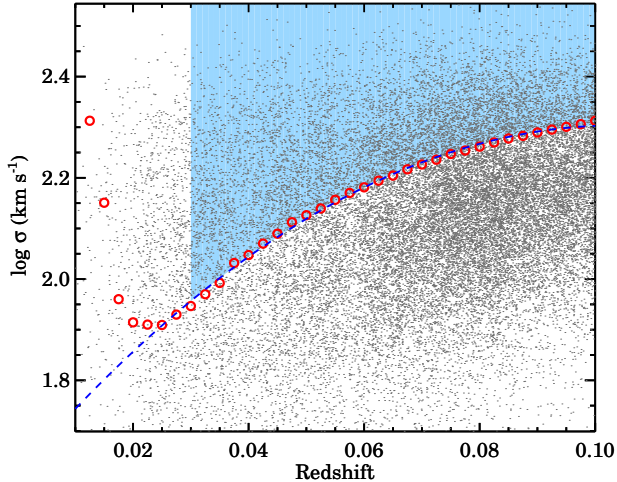}
\caption{Velocity dispersion limit (circles) as a function of redshift. 
The dashed line is a 2nd order polynomial fit to $\sigma_{lim} (z)$.
Only galaxies with $\sigma > \sigma_{lim} (z)$ (shaded region) 
 are included in estimating the VDF. 
We display 20\% of the data for clarity. }
\label{siglim}
\end{figure}

We derive $\sigma_{lim} (z)$ by repeating the $\sigma_{lim}$ determination 
 for a series of volume limited subsamples 
 with different $z_{max}$. 
For the series of volume limited samples, 
 $z_{max}$ ranges from 0.0125 to 0.10 in intervals of $\Delta z = 0.0025$.  
In these volume limited samples,  
 the $M_{r}$ limit is brighter as $z_{max}$ increases
 and thus $\sigma_{lim}$ increases. 
Red circles in Figure \ref{siglim} show 
 $\sigma_{lim}$ for the series of volume limited samples. 
The value of $\sigma_{lim}$ changes rapidly for $z < 0.03$ 
 mainly due to the small number of galaxies in the volume. 
We exclude these local galaxies from further analysis.
In the redshift range $0.03 \leq z \leq 0.10$, 
 the samples are complete for $\log \sigma > \log \sigma_{lim} (z)$
 (the shaded area in Figure \ref{siglim}).
We fit $\sigma_{lim} (z)$ with a simple polynomial (blue curve)
\begin{equation}
\log \sigma_{lim} (z) = 1.62 + 13.18 z - 63.38 z^{2}. 
\label{siglim_eq}
\end{equation}

\begin{figure}
\centering
\includegraphics[scale=0.38]{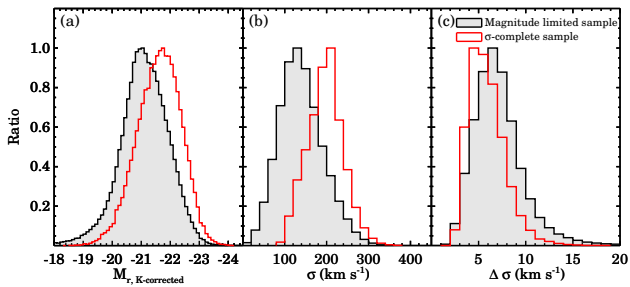}
\caption{Properties of galaxies in the magnitude limited sample (filled histogram)
 and in the $\sigma$ complete sample (open red histogram)
 including (a) $M_{r}$, 
 (b) $\sigma$ and (c) $\Delta \sigma$. }
\label{prop}
\end{figure}
 
\subsection{Velocity Dispersion Complete Sample}
We construct the $\sigma$ complete sample from the magnitude limited sample
 based on equation \ref{siglim_eq}. 
We select galaxies by taking only objects with $\sigma > \sigma_{lim}(z)$. 
The $\sigma$ complete sample consists of 40660 quiescent galaxies
 in the redshift range $0.03 \leq z \leq 0.10$. 
Figure \ref{prop} displays the $M_{r}$, $\sigma$ and 
 observational $\sigma$ uncertainty ($\Delta \sigma$) distributions 
 for galaxies in the $\sigma$ complete sample and in the magnitude limited sample. 
Galaxies in the $\sigma$ complete sample 
 are brighter and have smaller $\Delta \sigma$s 
 compared to the objects in the magnitude limited sample.
The typical $\Delta \sigma$s for galaxies 
 in the $\sigma$ complete sample are $< 10~\kms$. 
 
The $\sigma$ complete sample includes 18 galaxies with $\sigma > 400~\kms$, 
 extreme for field galaxies.
Among these, five objects have $\sigma$ measurements 
 with very large $\Delta \sigma$s ($\gtrsim 100~\kms$). 
Six objects show merging feature in the SDSS images. 
The merging feature may affect the central velocity dispersion measurements through the fiber. 
The remaining seven galaxies are bright elliptical galaxies 
 without signs of disks. 
These seven galaxies reside in known groups or clusters
 listed in the NASA Extragalactic Database (NED). 
 
The $\sigma$ complete sample 
 represents the general field sample and 
 includes galaxies in cluster regions
 according to the cluster abundance in the volume surveyed.
We estimate the fraction of cluster galaxies in the $\sigma$ complete sample 
 based on the CIRS cluster catalog \citep{Rines06}. 
The CIRS cluster catalog includes X-ray detected galaxy clusters 
 in the SDSS DR4 with a sky coverage of 4783 deg$^{2}$ 
 ($\sim60\%$ of the SDSS DR12). 
The CIRS catalog provides a well-defined sample of 
 spectroscopically identified galaxy clusters.

There are 64 CIRS clusters 
 in the redshift range of the $\sigma$ complete sample. 
We count the number of galaxies in the $\sigma$ complete sample 
 that are candidate members of the CIRS clusters. 
We identify the candidate members 
 with $R_{cl} < R_{200}$ and $|\Delta cz| / (1+z_{cl}) < 2000~\kms$, 
 where $R_{cl}$ is a distance of galaxy from the cluster center, 
 $R_{200}$ is the cluster radius at 200 times critical density,
 $z_{cl}$ is the cluster redshift.
The radial velocity difference limit
 reflects the maximum amplitude of the caustics for the CIRS clusters \citep{Rines06}. 
We find that 708 galaxies in the $\sigma$ complete sample 
 are possible members of the CIRS clusters. 
If the cluster abundance is similar in DR12, 
 the total number of galaxies possibly in cluster regions is $\sim1188$ objects
 (= $708 \times \Omega_{DR12} / \Omega_{DR4}$). 
Thus, the overall fraction of cluster galaxies in the $\sigma$ complete sample 
 is $\sim 3\%$. 
The contribution of cluster galaxies 
 increases at greater $\sigma$ 
 (e.g. it is $\sim5\%$ at $\sigma > 250~\kms$). 

\section{METHOD FOR CONSTRUCTING THE VELOCITY DISPERSION FUNCTION}\label{method}
 
There are several methods for measuring the 
 statistical distribution function of a galaxy property
 (i.e., luminosity, stellar mass and velocity dispersion functions)
 over a wide redshift range.
These methods include
 the classical 1/V$_{max}$ method \citep{Schmidt68}, 
 the parametric maximum likelihood method (\citealp{Sandage79}, STY), and 
 the non-parametric maximum likelihood method \citep{Efstathiou88}. 
Several studies review
 the impact of the different methods on the resulting shape of the luminosity and/or 
 the stellar mass functions 
 \citep{Willmer97, Takeuchi00, Weigel16}.

The statistical methods require different basic assumptions. 
For example, 
 results based on the 1/$V_{max}$ method can be biased 
 if large scale inhomogeneities dominate the sample.
Application of the STY method requires 
 a prior functional form for the distribution function. 

We employ the {\it stepwise maximum likelihood} (SWML) method 
 introduced by \citet{Efstathiou88}. 
This non-parametric method is powerful 
 because it is less biased in the presence of inhomogeneities in galaxy distribution 
 than the 1/V$_{max}$ method. 
The SWML method also has the advantage that 
 a prior functional form of the velocity dispersion function is unnecessary.

The velocity dispersion function derived according to the SWML method 
 can be described as \citep{Efstathiou88, Takeuchi00, Weigel16}
\begin{multline}
\Phi_{k} d\log \sigma = \sum^{N_{gal}}_{i=1} W(\log \sigma_{k} - \log \sigma_{i}) \\
 \times \bigg[ \sum^{N_{gal}}_{i=1} \frac{H(\log \sigma_{k} - \log \sigma_{lim, i})}{\sum^{N_{bin}}_{j=1} \Phi_{j} d\log \sigma H(\log \sigma_{j} - \log \sigma_{lim, i})} \bigg]^{-1}, 
\end{multline}
 where $W(x)$ and $H(x)$ are step functions defined as 
\begin{equation}
 W(x) =
  \begin{cases}
     1 & \text{for}~-\frac{\Delta \log \sigma}{2} \leq x \leq \frac{\Delta \log \sigma}{2}\\
     0 & \text{otherwise},
  \end{cases}
\end{equation}
\begin{equation}
 H(x) = 
  \begin{cases}
   0 & \text{for}~x < -\frac{\Delta \log \sigma}{2} \\
   \frac{x}{\Delta \log \sigma} + \frac{1}{2} 
      & \text{for} ~-\frac{\Delta \log \sigma}{2} \leq x \leq \frac{\Delta \log \sigma}{2} \\
   1 & \text{for}~x > \frac{\Delta \log \sigma}{2}.
  \end{cases}
\end{equation}
Here, $N_{gal}$ is the total number of galaxies in the sample, 
 $N_{bin}$ is the number of $\sigma$ bins, 
 $\sigma_{j}$ is the central $\sigma$ of the $\sigma$ bin $j$ and 
 $\sigma_{lim,i}$ is the minimum $\sigma$ that the $i$th galaxy 
 can have and still be found in the sample (see Table \ref{term} for summary).
We calculate $\sigma_{lim,i}$ for individual galaxies using equation \ref{siglim_eq}. 

Following \citet{Efstathiou88}, we also apply a constraint
\begin{equation}
g = \sum_{k} \Phi_{k} \Delta \log \sigma (\log \sigma_{k} - \log \sigma_{f})^{\beta} - 1 = 0,
\end{equation}
where $\log \sigma_{k}$ is the log of the mean $\sigma$ for each bin and 
 $\log \sigma_{f}$ is the log of a fiducial $\sigma$ and 
 $\beta$ is a constant. 
We choose $\beta = 1.5$ following \citet{Efstathiou88} and set 
 $\log \sigma_{f} = 1.0$ following \citet{Weigel16} who set
 a small fiducial stellar mass when they measure the stellar mass function 
 based on the SWML method. 
We derive the errors from the covariance matrix following \citet{Efstathiou88}
 (more details are in \citealp{Ilbert05, Weigel16}).  
We take the normalization $\Phi_{*}$ of the VDF
 following \citet{Takeuchi00} and \citet{Weigel16}
\begin{equation}
\Phi_{*} = \frac{1}{V_{\rm total}} \sum^{N_{gal}}_{i} \frac{1}{\int^{\infty}_{\log \sigma_{lim,i}} \Phi~d\log\sigma}. 
\end{equation} 
Here, $V_{\rm total}$ is the volume of the sample
\begin{equation}
V_{\rm total} = \frac{4 \pi}{3} \frac{\Omega_{\rm survey}}{\Omega_{\rm all~sky}} \bigg[d_{c} (z'_{max})^{3} - d_{c} (z'_{min})^{3}\bigg].
\end{equation}
Here, $\Omega_{\rm survey}~(= 8032~{\rm deg}^{2})$ is the solid angle of the SDSS survey,
 $\Omega_{\rm all~sky}~(= 40253~{\rm deg}^{2})$ is the solid angle of the full sky and 
 $d_{c} (z)$ is a comoving distance at redshift $z$. 
$z'_{min}~(= 0.03)$ and $z'_{max}~(= 0.10)$ are 
 the lower and upper redshift limit of the $\sigma$ complete sample, respectively. 

\begin{figure*}
\centering
\includegraphics[scale=0.95]{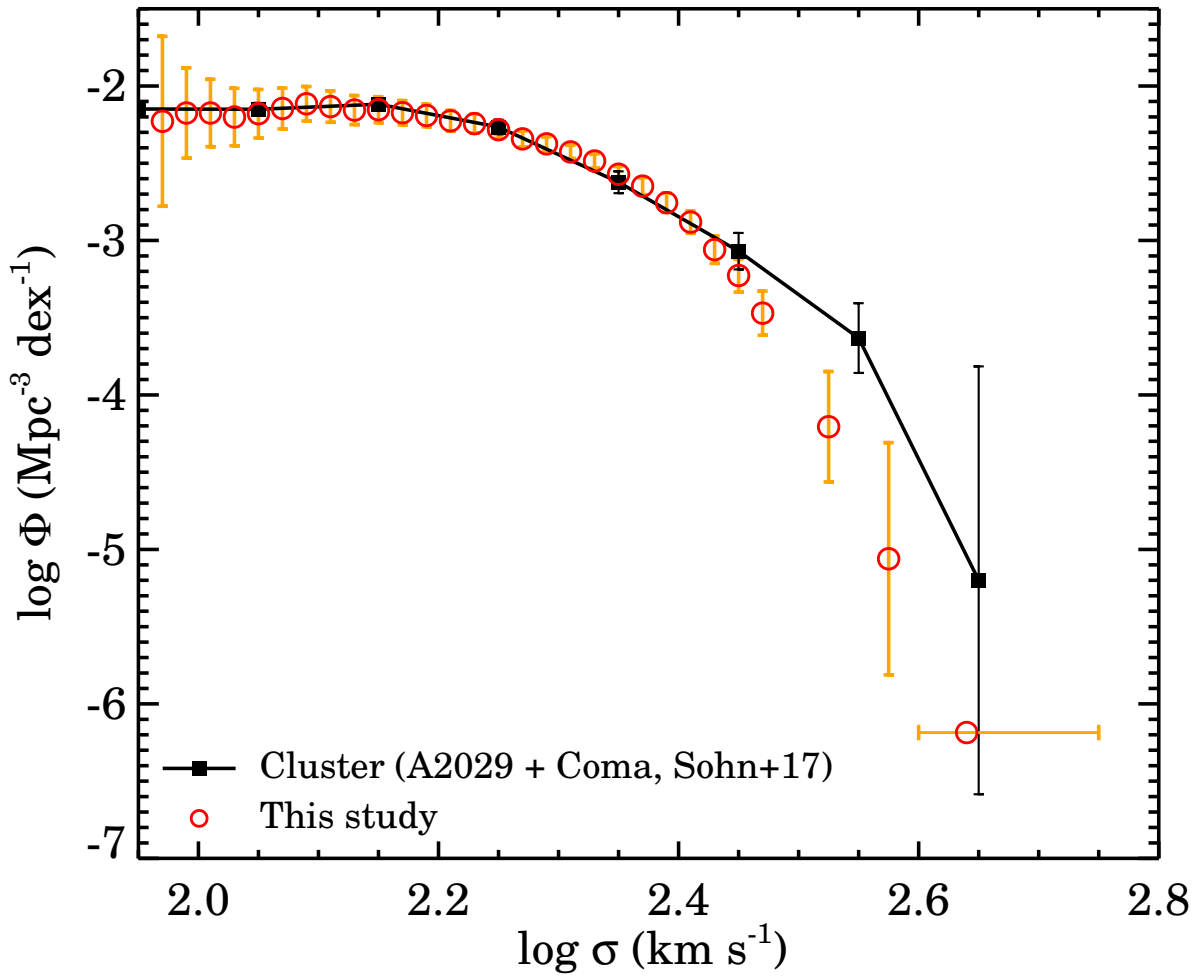}
\caption{VDF for the $\sigma$ complete sample (red circles)
 compared with the cluster VDF (black squares, the sum of Coma and A2029 VDFs, \citealp{Sohn17}). 
We arbitrarily scale the cluster VDF
 to compare the shapes of the two functions. }
\label{cl}
\end{figure*}

Figure \ref{cl} shows the VDF 
 for the $\sigma$ complete sample; the field VDF.
Table \ref{tab} lists the data points for the field VDF. 
The most massive point marks the seven galaxies with $\sigma > 400~\kms$.
The horizontal error bar displays 
 the minimum and maximum $\sigma$ for these galaxies. 
We plot the cluster VDF from \citet{Sohn17} for comparison. 
\citet{Sohn17} measure the cluster VDF of Coma and A2029 
 using the spectroscopically identified members
 within $R_{cl} < R_{200}~(\sim 2.0$ Mpc). 
They use the same quiescent galaxy selection $\dn > 1.5$ and 
 $\sigma$s corrected to the fiducial aperture 3 kpc. 
The VDFs for Coma and A2029 
 are essentially identical over the range $2.0 < \log \sigma < 2.6$
 (Figure 15 in \citealp{Sohn17}). 
 
\begin{deluxetable}{ccc}
\tablecolumns{3}
\tabletypesize{\scriptsize}
\tablewidth{0pt}
\tablecaption{Field Velocity Dispersion Function}
\tablehead{\colhead{$\log \sigma$} & \colhead{$\Phi (\Sigma)$} & \colhead{$\Phi$ error} \\
           ($\kms$) & (10$^{-3}$ Mpc$^{-3}$ dex$^{-1}$) & (10$^{-3}$ Mpc$^{-3}$ dex$^{-1}$)} 
\startdata
1.97 & 5.911 & 7.487 \\
1.99 & 6.687 & 4.500 \\
2.01 & 6.686 & 3.360 \\
2.03 & 6.297 & 2.696 \\
2.05 & 6.617 & 2.396 \\
2.07 & 7.152 & 2.179 \\
2.09 & 7.671 & 1.974 \\
2.11 & 7.358 & 1.722 \\
2.13 & 6.987 & 1.511 \\
2.15 & 7.024 & 1.360 \\
2.17 & 6.715 & 1.217 \\
2.19 & 6.445 & 1.077 \\
2.21 & 5.954 & 0.921 \\
2.23 & 5.709 & 0.803 \\
2.25 & 5.216 & 0.674 \\
2.27 & 4.547 & 0.545 \\
2.29 & 4.221 & 0.449 \\
2.31 & 3.743 & 0.371 \\
2.33 & 3.270 & 0.345 \\
2.35 & 2.685 & 0.313 \\
2.37 & 2.248 & 0.286 \\
2.39 & 1.757 & 0.253 \\
2.41 & 1.315 & 0.219 \\
2.43 & 0.872 & 0.178 \\
2.45 & 0.593 & 0.147 \\
2.47 & 0.338 & 0.111 \\
2.53 & 0.062 & 0.051 \\
2.58 & 0.009 & 0.015 \\
2.68 & 0.001 & 7.118 
\enddata
\label{tab}
\end{deluxetable}

\section{Discussion}

We compare the VDF we measure to the VDF derived from volume limited samples
 to understand systematic issues from the sample selection (Section \ref{dvlim}). 
We compare our result with previous field VDFs (Figure \ref{field}, Section \ref{sec_fld}) and 
 with the cluster VDF (Section \ref{sec_cl}). 
We comment on theoretical considerations (Section \ref{theo}). 

\subsection{$\sigma$ complete sample vs. volume limited sample}\label{dvlim}

A critical step in measuring the VDF
 is the use of a $\sigma$ complete sample. 
The $\sigma$ complete sample includes 
 every galaxy with $\sigma > \sigma_{lim} (z)$
 in the magnitude limited sample. 
We examine 
 the impact of the use of the $\sigma$ complete sample
 by comparing our measured VDF (Section \ref{method}) 
 with the result from a purely volume limited sample.  
    
\begin{figure}
\centering
\includegraphics[scale=0.49]{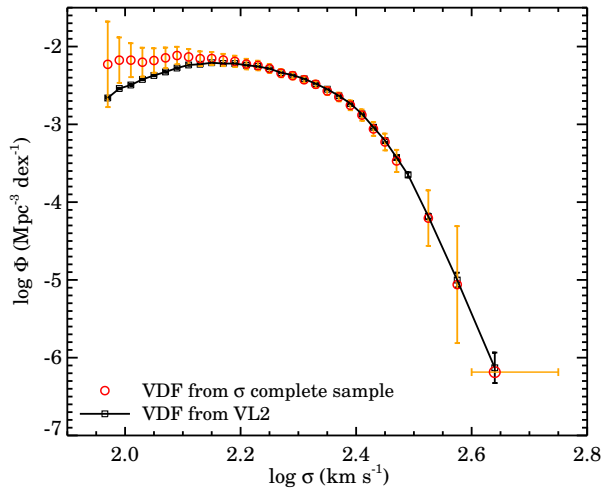}
\caption{VDFs for 
 the $\sigma$ complete sample (red circles) and 
 for VL2 (black squares). }
\label{vlim}
\end{figure}

Figure \ref{vlim} shows 
 the comparison between the $\sigma$ complete VDF and 
 a VDF derived from the volume limited sample, VL2. 
VL2 covers the redshift range $0.03 \leq z \leq 0.09$
 and $M_{r} \leq -20.49$ (Table \ref{sample}). 
The VDF for VL2 is the number of galaxies 
 in each $\sigma$ bin divided by the volume of the sample.
The VDFs for the $\sigma$ complete and the VL2 samples are 
 similar for $\log \sigma > 2.2$.
However, the VDF for the $\sigma$ complete sample remains flat 
 for $\log \sigma < 2.2$ whereas the VDF for VL2 declines
 as a result of incomplete sampling of the $\sigma$ distribution at every $M_{r}$. 

\begin{figure}
\centering
\includegraphics[scale=0.36]{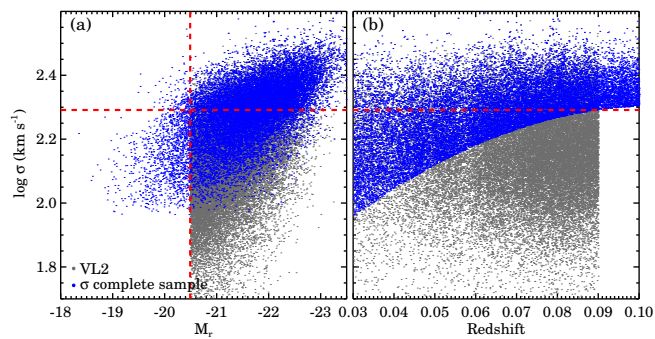}
\caption{
(a) $\sigma$ as a function of $M_{r}$ for VL2 (grey) and  
 for the $\sigma$ complete sample (blue), respectively. 
The vertical dashed line shows the magnitude limit ($M_{r, lim}$) of VL2. 
The objects in the $\sigma$ complete sample that are fainter than the magnitude limit 
 make the VDF flat at low $\sigma$ end. 
The horizontal dashed line displays the $\sigma$ completeness limit ($\sigma_{lim}$) 
 of VL2 (equation \ref{siglim_eq}). 
(b) $\sigma$ for the samples as a function of redshift. 
The symbols are the same as in the left panel.
We plot 20\% of the data for clarity.}
\label{comp_dvlim}
\end{figure}

VL2 is only complete for $\sigma > \sigma_{lim, VL2}~(= \log \sigma \gtrsim 2.3$),
 not for $\log \sigma \sim 2.0$. 
At $\log \sigma < 2.3$, 
 VL2 is incomplete 
 because galaxies fainter than $M_{r, lim}$ but with $\sigma$ in this range
 are removed from the sample. 
Figure \ref{comp_dvlim} (a) directly compares 
 the distribution of $\sigma$ for the VL2 sample and 
 the $\sigma$ complete sample as a function of $M_{r}$. 
VL2 does not completely sample the $\sigma$ distribution below $\sigma_{lim}$. 
VL2 is more incomplete at lower $\sigma$ 
 because more and more of these galaxies are fainter than $M_{r, lim}$. 
This incomplete sampling produces the downturn in the VDF for VL2. 
 
The VDF for VL2 appears comparable with the VDF for the $\sigma$ complete sample 
 to $\log \sigma \sim 2.2$ rather than to $\log \sigma \sim 2.3$,
 the $\sigma_{lim, VL2}$. 
Figure \ref{comp_dvlim} (b) shows $\sigma$ as a function of redshift 
 for VL2 and for the $\sigma$ complete sample. 
At $\log \sigma < 2.3$, 
 VL2 samples galaxies with $\sigma < \sigma_{lim} (z)$;
 these objects are excluded from the $\sigma$ complete sample.
These galaxies contribute to the VDF measurement based on VL2 and 
 make the VDF decline less rapidly. 
Thus, although the VDFs for VL2 and the $\sigma$ complete sample
 have similar shape to $\log \sigma \sim 2.2$, 
 the VDFs are actually derived from somewhat different galaxy samples. 
 
VDFs measured from the various volume limited samples 
 with different $z_{max}$ also decline in the low $\sigma$ regime, 
 but the downturns occur at different $\sigma$s.
Volume limited samples with lower $z_{max}$ 
 have fainter $M_{r, lim}$ and 
 are thus more complete to lower $\sigma$. 
For example, the VDF for a volume limited sample with $z_{max} = 0.04$ 
 appears complete to $\log \sigma \sim 2.0$.
This VDF for a local volume limited sample 
 also contains galaxies brighter than $M_{r, lim}$ 
 but with $\sigma < \sigma_{lim}$. 

\subsection{Comparison with Other Field VDFs}\label{sec_fld}

\begin{figure}
\centering
\includegraphics[scale=0.49]{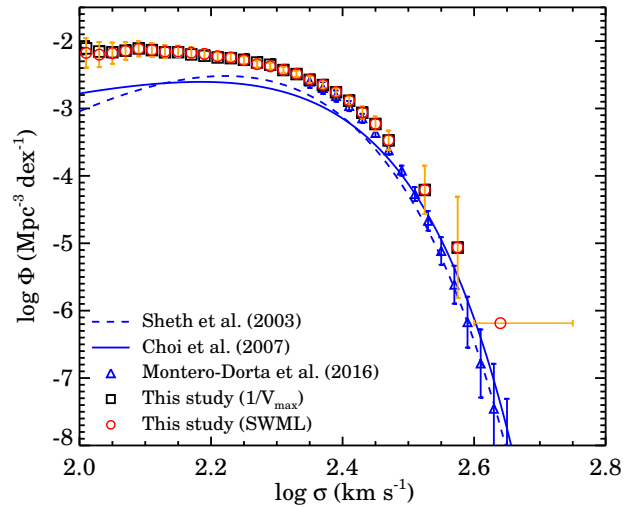}
\caption{Comparison between the field VDF
 for the $\sigma$ complete sample and previous field VDFs. 
Black squares and red circles display the field VDF 
 for the $\sigma$ complete sample derived with the $1/V_{max}$ and the SWML method,
 respectively. 
Dashed and solid lines show SDSS result \citep{Sheth03, Choi07}, and  
 blue triangles show the BOSS result \citep{Montero-Dorta17}, respectively. }
\label{field}
\end{figure}

Figure \ref{field} compares the $\sigma$ complete VDF 
 with previous field VDFs derived from SDSS \citep{Sheth03, Choi07} and
 BOSS \citep{Montero-Dorta17}. 
We plot the previous field VDFs 
 based on the fitting parameters for 
 the `modified' Schechter function (equation (4) in \citealp{Sheth03}) 
 given in the literature. 
We note that the BOSS VDF is limited to $\log \sigma > 2.35$ (blue squares). 
There are several other published field VDFs 
 including \citet{Mitchell05}, \citet{Chae10} and \citet{Bezanson11}. 
However, we compare our results only with other results 
 based on spectroscopic determinations of $\sigma$. 

The field VDFs in Figure \ref{field} are almost identical for $\log \sigma > 2.3$. 
The $\sigma$ complete VDF slightly exceeds the other SDSS VDFs; 
 this subtle discrepancy may result from different sample selection
 including the SDSS data release that we use and the difference in quiescent galaxy selection. 
The slopes of the VDFs are consistent in the high $\sigma$ region 
 regardless of the sample and the quiescent galaxy classification. 
The VDF measurement across samples and techniques is thus robust 
 for $\log \sigma > 2.3$. 

The most striking feature in the comparison 
 between field VDFs is 
 the significant difference for $\log \sigma < 2.3$. 
All previous field VDFs decline for $\log \sigma < 2.3$; 
 our field VDF remains flat. 
The slope for $\log \sigma < 2.3$ varies among the previous VDFs. 
The VDF from \citet{Choi07} is flatter than
 the VDF from \citet{Sheth03}. 
\citet{Choi07} suggest that 
 some morphologically identified early-type galaxies 
 in their sample that are not included 
 the spectroscopically identified early-type samples of \citet{Sheth03}
 possibly account for the flatter slope. 

Although a detailed analysis of the impact of sample selection 
 is beyond the scope of this paper, 
 a preliminary analysis suggests that 
 the form and parameters of the VDF actually have 
 little sensitivity to the sample selection. 
For example, 
 selection of quiescent galaxies with Sersic index $\geq 3$ yields a VDF 
 indistinguishable from the results contained here. 
A color selected red sequence also yields similar results. 
There may well be subtle dependencies on selection 
 which require a full detailed analysis. 
Our main goal here is to demonstrate that 
 the field and cluster VDF are indistinguishable 
 when the objects are selected in an identical manner.

The method for deriving the VDF may also affect the shape of the VDF. 
We measure the field VDF using the SWML method.
Previous studies use different methods. 
\citet{Sheth03} use the classical $1/V_{max}$ method 
 and \citet{Choi07} count the number of galaxies 
 in $\sigma$ bins divided by the survey volume. 

To examine the impact of different methods for deriving VDF, 
 we compute the VDF for the $\sigma$ complete sample 
 using the $1/V_{max}$ method. 
The resulting VDF is generally consistent with the VDF 
 derived from the SWML method. 
The only difference is that 
 the $1/V_{max}$ VDF slightly exceeds the SWML VDF at $\log \sigma < 2.05$ 
 (Figure \ref{field}). 
This result is consistent with the conclusion of \citet{Weigel16} 
 who measure the stellar mass functions using the $1/V_{max}$ and SWML methods. 
The stellar mass function derived by applying the $1/V_{max}$
 exceeds the SWML result at the low mass end. 
We conclude that the use of the SWML method cannot explain 
 the difference in the field VDFs for $\log \sigma < 2.3$. 
 
We measure the VDF based on the $\sigma$ complete sample; 
 previous VDFs are based on volume limited samples \citep{Choi07} or 
 magnitude limited samples after correction for incompleteness \citep{Sheth03, Montero-Dorta17}. 
The previous field VDFs show similar behavior
 to the VDFs we derive from purely volume limited samples in Section \ref{dvlim}. 
This result suggests that 
 the downturns in previous field VDFs may result from 
 incomplete sampling of the $\sigma$ distribution at every $M_{r}$.

\subsection{Comparison with Cluster VDFs}\label{sec_cl}

The field VDF we derive and the cluster VDF are
 based on identical sample selection. 
The field and cluster VDFs are both remarkably flat 
 for $\log \sigma < 2.3$ (Figure \ref{cl}). 
\citet{Sohn17} argue that the cluster VDF is a lower limit for $\log \sigma < 2.3$
 because low surface brightness objects are missing from the sample. 
The field VDF may also be affected by missing low surface brightness galaxies.

The cluster VDF exceeds the field VDF for $\log \sigma > 2.4$. 
Note that the field VDF also includes 
 a small fraction ($\sim3\%$) of cluster galaxies, 
 and the contribution of cluster galaxies is larger at greater $\sigma$. 
Thus, the gap between the cluster and the field VDFs would be even larger
 if we excluded cluster galaxies from our sample. 
\citet{Sohn17} first observed the excess in the cluster VDF. 
They suggest that the presence of BCGs
 and other very massive objects, 
 which are relatively rare in field samples, 
 produces this excess in the cluster VDF.

\begin{figure}
\centering
\includegraphics[scale=0.36]{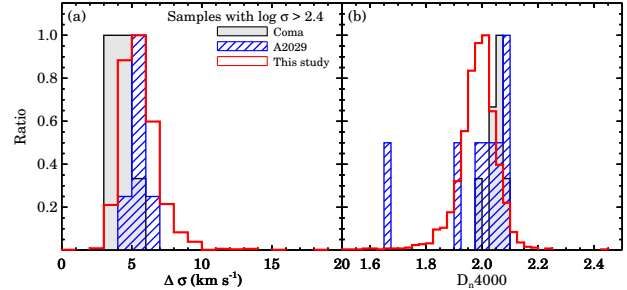}
\caption{(a) $\Delta \sigma$ distributions and (b) $\dn$ distributions 
 for galaxies with $\log \sigma > 2.4$.
The red open histogram shows the distributions for the $\sigma$ complete sample, 
 the filled and hatched histograms show the same for 
 Coma and A2029 members, respectively. }
\label{comp_cl}
\end{figure}

Figure \ref{comp_cl} (a) displays 
 the distribution of $\Delta \sigma$ for galaxies with $\log \sigma > 2.4$ 
 in both the field and the cluster samples. 
There is no significant difference 
 in the $\Delta \sigma$ distributions. 
Coma galaxies tend to have somewhat smaller $\Delta \sigma$s. 
Because of the proximity of Coma, 
 Coma galaxies at fixed $\sigma$ are generally brighter than 
 their counterparts in the A2029 and in the field samples. 
$\Delta \sigma$s for $\log \sigma > 2.4$ are $\lesssim 10~\kms$, 
 less than the size of $\sigma$ bin ($\sim 15~\kms$) we use for measuring the VDF. 
Thus, the shape of the VDFs and 
 the excess in the cluster VDF for $\log \sigma > 2.4$
 are not significantly affected by the error distribution. 
 
Figure \ref{comp_cl} (b) shows the $\dn$ distributions 
 for the field and the cluster samples. 
The cluster samples tend to have larger $\dn$
 (see also Figure 13 in \citealp{Sohn17}). 
The $\dn$ is the indicator of stellar population of ages and 
 the cluster objects tend to be older than the field objects. 
This result is consistent with the correlation 
 between $\sigma$ and $\dn$ \citep{Zahid17}. 
 
The excess of the cluster VDF at high $\sigma$ implies that 
 there may be a fundamental difference in the relative abundance of galaxies with $\log \sigma > 2.4$
 in the field and cluster environments.
The shallower cluster VDF implies that 
 the clusters preferentially contain a larger number of high $\sigma$ galaxies 
 than the typical field.
Each cluster has a dominant central galaxy (brightest cluster galaxy, BCG)
 generally with $\log \sigma > 2.4$ \citep{Lauer14}. 
However, the excess in the cluster VDF
 cannot be explained by a single BCG in each cluster. 
Fifteen galaxies with $\log \sigma > 2.4$ make
 the slope of the cluster VDF shallower at $\log \sigma > 2.4$.

Clusters often consist of several subclusters each with its own BCG. 
Their presence reflects the hierarchical formation process. 
High $\sigma$ galaxies that develop in the individual subclusters
 may contribute to the excess in the cluster VDF. 
For example, the multiple BCGs within several substructures in the Coma cluster 
 \citep{Colless96, Tempel17}
 may support this picture.
Analysis of cluster VDFs in concert with substructure analysis
 for a large set of clusters should further elucidate this issue.  
 
\subsection{Comparison with Simulations}\label{theo}

VDFs provide an interesting probe for modeling structure formation 
 because $\sigma$ may be a good proxy for the DM subhalo mass
 \citep{Wake12a, Wake12b, Bogdan15, Zahid16}. 
Here we compare VDFs for the field and for two massive clusters. 
The difference between the field and cluster VDFs for $\log \sigma > 2.4$ 
 is a potential test of models for the formation of galaxies in clusters.

Theoretical model VDFs have not yet been computed in a way that mimics the observations. 
\citet{Munari16} investigate the $\sigma$ distribution of a rich cluster A2142 
 based on $\sigma$s from the SDSS and compare their result with simulations. 
They suggest that numerical simulations significantly underestimate the number of massive subhalos. 
In fact, they compare the distribution of circular velocities of galaxies 
 in the observations and the simulations 
 because measuring circular velocities is straightforward in simulations. 
To derive the circular velocity distribution for the observations, 
 they convert the observed central velocity dispersions into circular velocities. 
This conversion makes reasonable assumptions, 
 but a direct comparison of central $\sigma$s would provide more powerful constraints. 

Currently there are no VDFs derived from simulations that mimic the observations. 
To do this, the velocity dispersions for subhalos in the simulations 
 must be measured in a cylinder through the center of the subhalo
 mimicking the $\sigma$ we measure observationally. 
Comparison with the observed field VDFs 
 requires model VDFs derived from hydrodynamic large-scale 
 cosmological simulations with high resolutions like 
 Illustris \citep{Vogelsberger14}, EAGLE \citep{Schaye15}, 
 and MassiveBlack-II \citep{Khandai15}. 
These simulations identify galaxies with stellar mass $> 10^{9} M_{\odot}$ and thus 
 the VDF from this simulations could in principle be measured for $\log \sigma > 2.0$. 
Several studies do calculate 
 the projected radial velocity dispersions for objects 
 in cosmological hydrodynamic simulations \citep{Choi17, Penoyre17}. 
Their velocity dispersion measurements mimic IFU observations, 
 but they do not provide $\sigma$ in fixed metric apertures.
Velocity dispersions measured within circular apertures could be derived from 
 these simulations and would be the best benchmark 
 for direct comparison with the observed VDFs. 
The combinations of observed and simulated VDFs are 
 a potentially important window for understanding the halo mass distribution. 

\section{CONCLUSION}

We measure the field VDF 
 for a $\sigma$ complete sample in the range $\log \sigma > 2.0$.
The $\sigma$ complete sample includes 
 field quiescent galaxies with $\dn > 1.5$ and $\sigma$ measured within a 3 kpc aperture, 
 just like the cluster sample used for measuring the cluster VDF \citep{Sohn17}.  
 
The $\sigma$ complete sample we construct includes 
 every galaxy with $\sigma$ larger than 
 the $\sigma$ completeness limit of the SDSS magnitude limited sample 
 as a function of redshift, $\sigma_{lim} (z)$. 
We estimate the $\sigma$ complete limit of the magnitude limited sample 
 empirically to construct the $\sigma$ complete sample.
The method we use to construct the $\sigma$ complete sample 
 can be applied more generally to construct 
 samples complete in any other observables from a magnitude limited survey. 

The VDF for the $\sigma$ complete sample is much flatter than 
 the VDF for purely volume limited sample for $\log \sigma < 2.2$. 
The $\sigma$ complete sample includes faint low $\sigma$ galaxies 
 below the $M_{r, lim}$ of the volume limited sample. 
The sampling of these galaxies results in the flatter VDF. 
Previous field VDFs derived from either a purely volume limited sample or a magnitude limited sample 
 also show a downturn for $\log \sigma < 2.2$.
The incompleteness of these samples in sampling the full $\sigma$ distribution
 could produce the decline in the previous VDFs at low $\sigma$.   
 
The flat field VDF we derive from the $\sigma$ complete sample 
 is essentially identical to the cluster VDF for $\log \sigma < 2.2$. 
The previous discrepancy in the low $\sigma$ range 
 appears to be completely explained by the use of the $\sigma$ complete sample.  
 
The field VDF is steeper than 
 the cluster VDF for the massive clusters A2029 and Coma
 for $\log \sigma > 2.4$. 
The difference for $\log \sigma > 2.4$ probes 
 the formation and evolution of massive galaxies in the cluster environment
 and may reflect the presence of these massive objects within cluster substructures. 
 
The VDF is potentially a direct probe of the dark matter subhalo mass distribution. 
The combination of the cluster VDF from \citet{Sohn17} and 
 the field VDFs provides benchmarks for testing N-body and hydrodynamic simulations. 
Measuring $\sigma$ from simulations in a way 
 that mimics the observations 
 would be an important step toward realizing the potential of this test.

\acknowledgments
We thank the referee for a clear report that improved the clarity of the paper. 
J.S. gratefully acknowledges the support of the CfA Fellowship. 
The Smithsonian Institution supported research of M.J.G.
H.J.Z is supported by the Clay Postdoctoral Fellowship. 
This research has made use of NASA's 
 Astrophysics Data System Bibliographic Services. 
Funding for SDSS-III has been provided by the Alfred P. Sloan Foundation, 
 the Participating Institutions, the National Science Foundation, 
 and the U.S. Department of Energy Office of Science. 
The SDSS-III web site is http://www. sdss3.org/. 
SDSS-III is managed by the Astrophysical Research Consortium for 
 the Participating Institutions of the SDSS-III Collaboration 
 including the University of Arizona, 
 the Brazilian Participation Group, Brookhaven National Laboratory, 
 University of Cambridge, Carnegie Mellon University, 
 University of Florida, the French Participation Group, 
 the German Participation Group, Harvard University, 
 the Instituto de Astrofisica de Canarias, 
 the Michigan State/Notre Dame/ JINA Participation Group, 
 Johns Hopkins University, Lawrence Berkeley National Laboratory, 
 Max Planck Institute for Astrophysics, 
 Max Planck Institute for Extraterrestrial Physics, 
 New Mexico State University, New York University, Ohio State University, 
 Pennsylvania State University, University of Portsmouth, Princeton University, 
 the Spanish Participation Group, University of Tokyo, University of Utah, 
 Vanderbilt University, University of Virginia, University of Washington, and Yale University.





\end{document}